\documentclass{aa}
\usepackage{psfig}
\usepackage{latexsym}

\newcommand{\gtap}{\mathrel{\hbox{\rlap{\lower.55ex \hbox {$\sim$}}
                   \kern-.3em \raise.4ex \hbox{$>$}}}}
\newcommand{\ltap}{\mathrel{\hbox{\rlap{\lower.55ex \hbox {$\sim$}}
                   \kern-.3em \raise.4ex \hbox{$<$}}}}
\newcommand{\cmsq}{{\rm cm}^{-2}}
\newcommand{\cts}{{\rm cts}\,{\rm s}^{-1}}
\newcommand{\ergs}{{\rm erg}\,{\rm s}^{-1}}

\begin{document}
\thesaurus{08.09.2; 08.14.1; 10.07.3; 13.25.5}
\title{X-ray and optical follow-up observations of the August 1998 X-ray 
transient in NGC\,6440}
\author{F. Verbunt\inst{1} \and M.H. van Kerkwijk\inst{1} \and
J.J.M. in 't Zand\inst{2} \and J. Heise\inst{2} 
}
\offprints{Frank Verbunt}
\mail{verbunt@phys.uu.nl}

\institute{   Astronomical Institute,
              P.O.Box 80000, 3508 TA Utrecht, the Netherlands
         \and Space Research Organization Netherlands, 
              Sorbonnelaan 2, 3584 CA Utrecht, the Netherlands
                }
\date{Received 2000 March 14; revised May 1; accepted May 15}   
\authorrunning{F.\ Verbunt et.\ al.}
\titlerunning{X-ray and optical observations of the X-ray 
transient in NGC\,6440}
\maketitle


\begin{abstract}
The BeppoSAX Wide Field Cameras detected a transient in
NGC\,6440 on 1998 Aug 22. ROSAT did not detect this source on 1998 Sep 8, 
indicating that the X-ray flux had decreased by a factor 400 at least,
and/or that the X-ray spectrum had become appreciably softer.
Analysis of archival ROSAT HRI data reveals two sources associated with
NGC\,6440; one of these may be the transient in quiescence.
We have also obtained B and R images of NGC\,6440 on 1998 Aug 26
and again on 1999 July 15, when the transient had returned
to quiescence. Subtraction of these images reveals one object in the core
which was brighter in B, but not in R, during the X-ray outburst.
We estimate $B\simeq22.7$ and $(B-R)_0\ltap0$ on 1998 Aug 26, which makes it
a viable candidate counterpart.
\keywords{Stars: individual: SAX J1748.9-2021, MX 1746-20 - stars: neutron
- globular clusters: individual: NGC6440 - X-rays: stars
}
\end{abstract}

\section{Introduction}

NGC\,6440 is a globular cluster near the center of the Galaxy,
at a distance of 8.5\,kpc and reddened by $E(B-V)=1.0$ 
(Ortolani et al.\ 1994).
A bright X-ray source was detected near this cluster with OSO-7 and with 
UHURU from 1971 December 17 to 1972 January 21
(Markert et al.\ 1975, Forman et al.\ 1976).
UHURU observations obtained before 1971 Oct 23 and 
after 1972 Mar 1 did not detect the source (Forman et al.\ 1976).
During the outburst the transient X-ray source had a virtually
constant luminosity of about $3\times 10^{37}\ergs$, in the 2-11\,keV band.
Before and after the outburst the flux was less than 5\%\ of this.
(We use the conversion of UHURU $\cts$ to flux given by Bradt \&\ 
McClintock 1983; and the absorption column $N_H=6.9\times10^{21}\cmsq$
determined by in 't Zand et al.\ 1999.)
\nocite{obb94}\nocite{mbc+75}\nocite{fjt76}\nocite{bm83}\nocite{zvs+99}

A dim source was detected in the core of NGC\,6440 with the Einstein
satellite, and again with ROSAT, at a luminosity of
$\sim10^{33}\ergs$, in the 0.5-2.5\,keV band (Hertz \&\ Grindlay 1983; 
Johnston et al.\ 1995); after conversion to the 2-11\,keV band,
this corresponds to $\ltap 10^{-4}$ of the outburst flux.
\nocite{hg83}\nocite{jvh95}

On 1998 August 22 a bright transient source appeared again in NGC\,6440,
observed with BeppoSAX. 
The position coincides with the globular
cluster within the accuracy of 1$'$.
This time the outburst lasted rather shorter: the source had a luminosity
of $6.0\times 10^{36}\ergs$ on Aug 22, $3.6\times 10^{36}\ergs$ on Aug 26,
and $< 10^{36}\ergs$ on Sep 1, in the 2-10\,keV band (in 't Zand et al.\ 1999,
also see Fig.\,\ref{xcur}).

Like persistent bright X-ray sources, transients occur more often in globular
clusters per unit of stellar mass than in the galactic disk.
To understand this overabundance one would like to study these sources
at optical and ultraviolet wavelengths.
So far, no transient X-ray source in a globular cluster has been optically
identified. Such identification is difficult
because the relatively large error circle of the X-ray position contains a
large number of stars. We therefore obtained a ROSAT HRI observation
as soon as possible after the detection of the transient with BeppoSAX,
in the hope of improving the X-ray position.
The optical brightness of soft X-ray transients is known to vary
in tandem with the X-ray luminosity (for a review, see e.g.\ Chen et 
al.\ 1997).
We therefore obtained optical images of NGC\,6440 to look for objects
that vary in tandem with the X-ray flux, in the hope of identifying the
optical counterpart of the transient.
\nocite{csl97}

In Sect.\,2 we describe the results of the new ROSAT HRI observation, and also 
analyse archival ROSAT data of NGC\,6440. In Sect.\,3 we describe the optical
observations and the search for an optical counterpart to the X-ray source.
Our results and their implications are discussed in Sect.\,4.

\section{X-ray observations}

\begin{table}
\caption{Log of the ROSAT HRI observations of NGC\,6440.
For each observation we give the JD at start and end of the exposures and the 
effective exposure time, and the number of counts and the countrate for
the central source, according to the standard reduction. Errors for detections
are 1-$\sigma$; the upper limits are 2-$\sigma$.
\label{tablog}}
\begin{tabular}{@{}llrc@{\hspace{0.2cm}}c}
date & exposure & $t_{\rm exp}$ & counts & countrate  \\
      &  start \&\ end & (s)        &   & (cts/ksec) \\
1991 Mar & 2448335.100-36.796 & 5779 & 10$\pm$3 & 1.7$\pm$0.6 \\
1992 Sep & 2448882.685-82.705 & 1522 &  $<$8 & $<$5.3  \\
1993 Mar & 2449066.507-68.662 & 27056 & 79$\pm$10 & 2.9$\pm$0.4 \\
1994 Mar & 2449425.378-34.410 & 2769 &  $<$8 & $<$2.9  \\
1994 Sep & 2449611.228-13.774 & 15172 & 38$\pm$7 & 2.5$\pm$0.5 \\
1998 Sep & 2451064.743-64.766 & 1940 & $<$10 & $<$5.2 \\
\end{tabular}
\end{table}

The X-ray observations were obtained with the ROSAT X-Ray Telescope
(Tr\"umper et al.\ 1991) in combination with the High-Resolution Imager
(HRI, David et al.\ 1995).
\nocite{tha+91}\nocite{dhkz95}
The log of the observations is given in Table~\ref{tablog}; the last entry 
in that table is the one obtained near the BeppoSAX observations, the other 
entries refer to earlier observations in the ROSAT data archive.
The standard data reduction was done with the Extended Scientific Analysis 
System (Zimmermann et al.\ 1996).
\nocite{zbb+96}
To take into account the re-calibration of the pixel size (Hasinger et 
al.\ 1998), we multiply the $x,y$ pixel coordinates of each photon with
respect to the HRI center with 0.9972. 
Then a search for sources is made by comparing counts in a box with
the counts in a ring surrounding it, and by moving this detection
box across the image.
The sources thus detected are excised from the image and a background map
is made for the remaining photons.
A search for sources is made by comparing the number of photons in
a moving box with respect to the number expected on the basis of the
background map.
Finally, at each position in which a source was found, 
a maximum-likelihood technique is used to compare the observed
photon distribution with the point spread function of the HRI
(Cruddace et al.\ 1988).
\nocite{hbg+98}\nocite{chs88}
The resulting countrates for significant detections
near the centre of each image are given in Table~\ref{tablog}. 

\subsection{The 1998 observation}

No source is detected in the 1998 Sep 8 observation. For a point source at
the center of the image, 90\%\ of the photons arrive within a circle
with a 5$''$ radius, in stable HRI pointings (David et al.\ 1995). 
At the time of observation the ROSAT satellite pointing was experiencing 
difficulties, effectively extending the radius of the point spread function 
by a few arcseconds. We therefore search
a circle with a 10$''$ radius around the center of NGC\,6440
(according to Picard \&\ Johnston 1995; see Table\,\ref{tabpos}), 
only five photons are detected.
The maximum of five photons detected remains if we
move the center of the circle to any location within 30$''$ of the
nominal cluster centre, thus allowing for possible inaccurate reconstruction
of the satellite pointing. 
\nocite{dhkz95}\nocite{pj95}

For an expected number of 10 photons, the Poisson probability of detecting 5
or fewer photons is 7\%.  We thus take 10 photons as the 2-$\sigma$
upper limit, which for the exposure of 1940\,s gives an upper limit
for the count rate of 0.005\,$\cts$.  To convert this countrate into a
luminosity we use one of the fits made to the BeppoSAX data by in 't
Zand et al.\ (1999), viz.\ a sum of a black body with temperature
$kT=0.84$\,keV and a bremsstrahlung spectrum with temperature
$46.6$\,keV, absorbed by a column $N_H=6.9\times10^{21}\cmsq$.  
In the ROSAT bandpass of 0.5-2.5\,keV the bremsstrahlung and blackbody
components contribute 83 and 17 \%, respectively, to the total flux.
For this spectrum, the upper limit of 0.005\,$\cts$ in the HRI corresponds
to an X-ray luminosity of $6\times 10^{33}\ergs$ between 0.5 and 2.5 keV,
or $1.5\times 10^{34}\ergs$ between 2 and 10 keV. 
Note that
the ROSAT range from 0.1-2.5\,keV is effectively limited to
above 0.5\,keV because of the high reddening. Since SAX measures the flux
down to 2 keV, the estimates of the ROSAT flux are quite accurate.

This implies that the flux of the transient in NGC\,6440 dropped by a
factor 250 or more between the BeppoSAX observation on Aug 26 and the ROSAT
HRI observation on Sep 8.

\begin{figure}
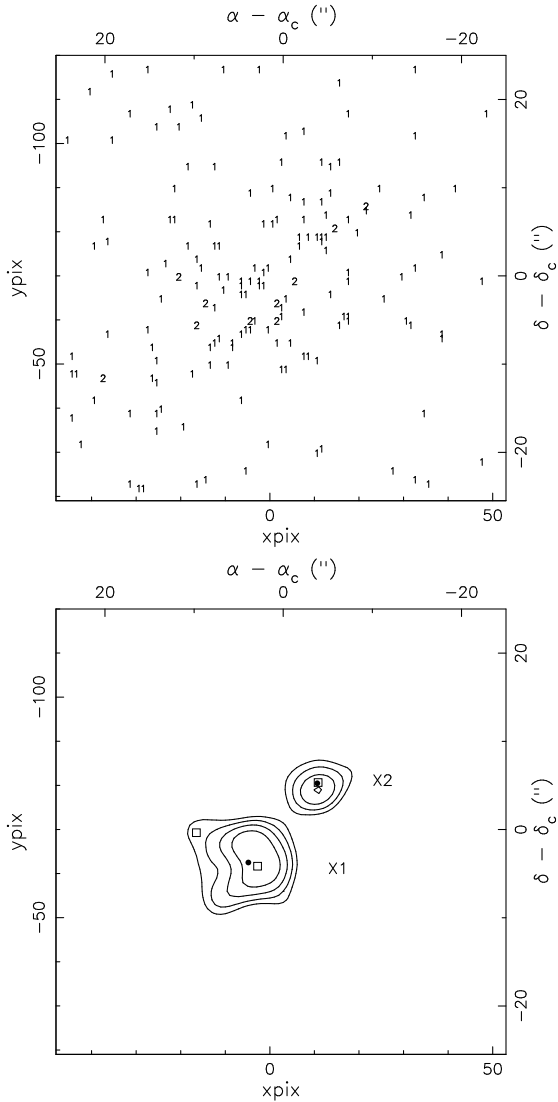


\centerline{\psfig{figure=figa.ps,width=0.85\columnwidth,clip=t} {\hfil}}

\centerline{\psfig{figure=figb.ps,width=0.85\columnwidth,clip=t} {\hfil}}

\caption{The central area of NGC\,6440 as observed with the ROSAT HRI
in 1993 March. The top frame shows the locations on the detector of the 
detected photons. 
176 photons fall within the frame, of which about 70 are
from the sources.
The lower frame the X-ray contours obtained after smoothing
with a 2-d Gaussian of width $\sigma=2''$.
The positions of the sources in a fit with two central sources
are indicated with $\bullet$; the positions for a fit with three sources
are indicated $\Box$. 
\label{figml}}
\end{figure}

\subsection{Earlier ROSAT observations: standard analysis}

The ROSAT data archive contains several hitherto
unpublished observations of NGC\,6440 made with the ROSAT HRI
after the 1991 observation reported by Johnston et al.\ (1995). 
A list of all ROSAT observations is given in Table\,\ref{tablog}.
We have analyzed each observation separately with the standard procedure,
and detect the source in NGC\,6440 in the 1993 observation and in the
Sep 1994 observation, i.e.\ in the observations with the longer exposure
times. In the shorter observations, we only obtain upper limits.
From the observed number of 3 counts in a circle with 5$''$ radius
near the cluster center we derive an upper limit
to the countrate of 8 counts for the central source for both
the 1992 and the March 1994 observation.
The three detections are compatible with a constant countrate, at a
level below the derived upper limits. We therefore have no indication
of variability between the ROSAT observations. On the other hand,
a variation by a factor 2 in the countrate between the 1991 and the
1993 detections is well within the range allowed by the
limited statistics.

The only source, other than the central source, detected significantly
in any of the ROSAT HRI pointings, is a point source in the 1993 observation,
listed as X3 in Table\,\ref{tabpos}.
There is no bright ($V\ltap14$) optical counterpart to this source in the 
digitized sky survey; and no object in SIMBAD within 1$'$ of its position.
We therefore cannot determine an accurate bore sight correction, and 
the uncertainty in the position of the central X-ray source is dominated
by the inaccuracy of the bore sight determination, which is about 5$''$
(1-$\sigma$; see David et al.\ 1995).
With this uncertainty, the three positions found for the central source
in 1991, 1993 and 1994 Sep are all compatible.

For each of the two longest observations, the standard analysis indicates
that the central source is extended, i.e.\ the distributions of
the photons is not compatible with that of a single point source.
No such indication is present in the 1991 observation, which has
a very small number of detected counts.

\subsection{Closer investigation of the central source}

To establish the nature of the extension of the central source
we first study the longest observation, obtained in 1993.
A smoothed X-ray image of the central region of NGC\,6440, shown in 
Fig.\,\ref{figml},
suggests that two sources are present. They are too close to be separated
by the standard analysis. We therefore implement a further analysis,
based on the maximum-likelihood method (see e.g.\ Cash 1979, Mattox et
al.\ 1996), as follows.
\nocite{cas79}\nocite{mbc+96}

The probability at detector pixel $i$ to obtain $n_i$ photons when a model
predicts $m_i$ photons is described by Poisson statistics
\begin{equation}
P_{i} = {{m_i}^{n_i}e^{-m_i}\over n_i!}
\end{equation}
The probability that the model describes the observations is
given by the product of the probabilities for all $i$ in the region considered:
$L'=\Pi P_i$. 
For computational ease we maximize the logarithm of this quantity:
\begin{equation}
\ln L' \equiv \sum_{i} \ln P_{i} = \sum_{i}n_i\ln m_i -\sum_{i}m_i 
 -\sum_{i}\ln n_i!
\end{equation}
The last term in this equation doesn't depend on the assumed model,
and -- in terms of selecting the best model -- may be considered as a
constant. Thus maximizing $L'$ is equivalent to minimizing $L$, where 
\begin{equation}
\ln L \equiv -2(\sum_{i}n_i\ln m_i -\sum_{i}m_i )
\end{equation}

Our further analysis of the 1993 ROSAT HRI observation is limited to the
central $50''\times50''$ area, and consists of four steps, in which
we fit a constant background, or a constant background plus one, two,
or three sources.
The values of $\ln L$ for the best models of these fits are
denoted as $\ln L_0$,  $\ln L_1$,  $\ln L_2$, and $\ln L_3$ respectively.
The significance of the $n$th source is found by comparing
$\ln L_{n-1}-\ln L_n$ with a $\chi^2$ distribution for three degrees of freedom
(for the flux and position of the source).

\begin{table}
\caption{Results of further maximum likelihood analysis of the
ROSAT HRI observations of two central sources in NGC\,6440.
For each observation we give the number of counts and the countrate for
both sources X1 and X2, the significance of detection (in terms of
$\Delta\chi^2\equiv\ln L_0-\ln L_1$ for X1, $\equiv\ln L_1-\ln L_2$ for X2) 
and the difference in position between the two sources
$\Delta\alpha\equiv\alpha_2-\alpha_1$ and 
$\Delta\delta\equiv\delta_2-\delta_1$.
\label{tabml}}
\begin{tabular}{lrrrrrrrr}
date & \multicolumn{2}{c}{counts} & \multicolumn{2}{c}{cts/ksec} &
\multicolumn{2}{c}{$\Delta\chi^2$(3dof)} & $\Delta\alpha$ & 
$\Delta\delta$ \\
         & X1 & X2 & X1  & X2  & X1 & X2 & $''$ & $''$ \\
1991 Mar & 10 &  4 & 1.8 & 0.8 & 29 & 10 & $-$9.3 & 8.5 \\
1993 Mar & 44 & 24 & 1.6 & 0.9 & 76 & 39 & $-$7.7 & 9.0 \\
1994 Sep & 15 & 14 & 1.0 & 0.9 & 22 & 19 & $-$7.2 & 8.5 \\
\end{tabular}
\end{table}

In performing the fits, we use the analytical result that the best fit has
a number of model photons equal to the number of detected photons.
Thus, in the model of a constant background, i.e.\ a constant value
of $m_i$, the optimum value of $m_i$ is found directly by dividing the
observed number of photons by the number of pixels under consideration.

In fitting a constant plus one source, we distribute the source
counts around the source position according to the ROSAT HRI point spread 
function at the center of the detector (David et al.\ 1995).
We then vary the background and source counts, and the source
position, to minimize $\ln L$. In doing so we keep the sum of the
source and background counts fixed at the observed number.

Next, we fit a constant background plus two or three point sources.
In this fit, the parameters of the first source are allowed to vary;
we use a Downhill Simplex method as implemented by Press et al.\ (1992)
to minimize $\ln L$ with respect to the 7 or 10 variables.
Here again, the sum of the model counts for the background and the
two/three sources is kept constant, at the observed number.
\nocite{ptvf92}

The results of this fitting procedure are summarized in Table\,\ref{tabml}; the
resulting source positions are shown in Fig.\,\ref{figml} and listed
in Table~\ref{tabpos}. A third source in the center is nominally significant 
at the 3-$\sigma$ level; however, our analysis doesn't take into
account any remaining jitter in the Point Spread Function, and we consider
the existence of this source not proven. Therefore, we do not list the source
in Table\,\ref{tabpos}; its position is shown in Fig.\,\ref{figml}.

\begin{table}
\caption{Positions of the SAX transient, of the
three X-ray sources detected in the 1993 March ROSAT
observation, of the center of NGC\,6440 (Picard \&\ Johnston 1995), and of  
optical variables
discussed in the text. X1 and X2 are in the center of NGC\,6440; X3 is
not related to the cluster. The errors $\Delta$ in the X-ray positions 
refer only to the statistical error; the overall positional error is 
dominated by the uncertainty of 5$''$ in the projection of the HRI on the sky. 
We also give the countrates, and for the sources in the cluster the
luminosity in the 0.5-2.5\,keV band.
\label{tabpos}}
\begin{tabular}{@{}l@{~~}r@{ }r@{ }r@{ }r@{ }r@{ }rccl}
 &  \multicolumn{3}{c}{$\alpha$\,(2000)}
   &  \multicolumn{3}{c}{$\delta$\,(2000)} & $\Delta$ & cts/ksec &
 $L_{\rm x}$\,($\ergs$) \\
SAX & 17 & 48 & 53.4\phantom{0} & $-$20 & 21 & 43\phantom{.0} & 60$''$ \\
X1 & 17 & 48 & 52.98 & $-$20 & 21 & 40.6 & 1$\farcs$0 & 1.6$\pm$0.3 &
 $2.0\times10^{33}$ \\
X2 & 17 & 48 & 52.43 & $-$20 & 21 & 31.7 & 1$\farcs$2 & 0.9$\pm$0.3 &
 $1.1\times10^{33}$ \\
X3 & 17 & 48 & 34.91 & $-$20 & 25 & 46.0 & 1$\farcs$2 & 0.9$\pm$0.2\\
C  & 17 & 48 & 52.7\phantom{0}  & $-$20 & 21 & 36.9 & 1$\farcs$0 \\
V0 & 17 & 48 & 52.40 & $-$20 & 21 & 38.7 & 0$\farcs$5 \\
V1 & 17 & 48 & 52.62 & $-$20 & 21 & 39.5 & 0$\farcs$5 \\
V2 & 17 & 48 & 52.14 & $-$20 & 21 & 32.6 & 0$\farcs$5 \\
\end{tabular}
\end{table}
\nocite{lmd96}\nocite{pj95}

\section{Optical observations}

\begin{figure*}
\centerline{\psfig{figure=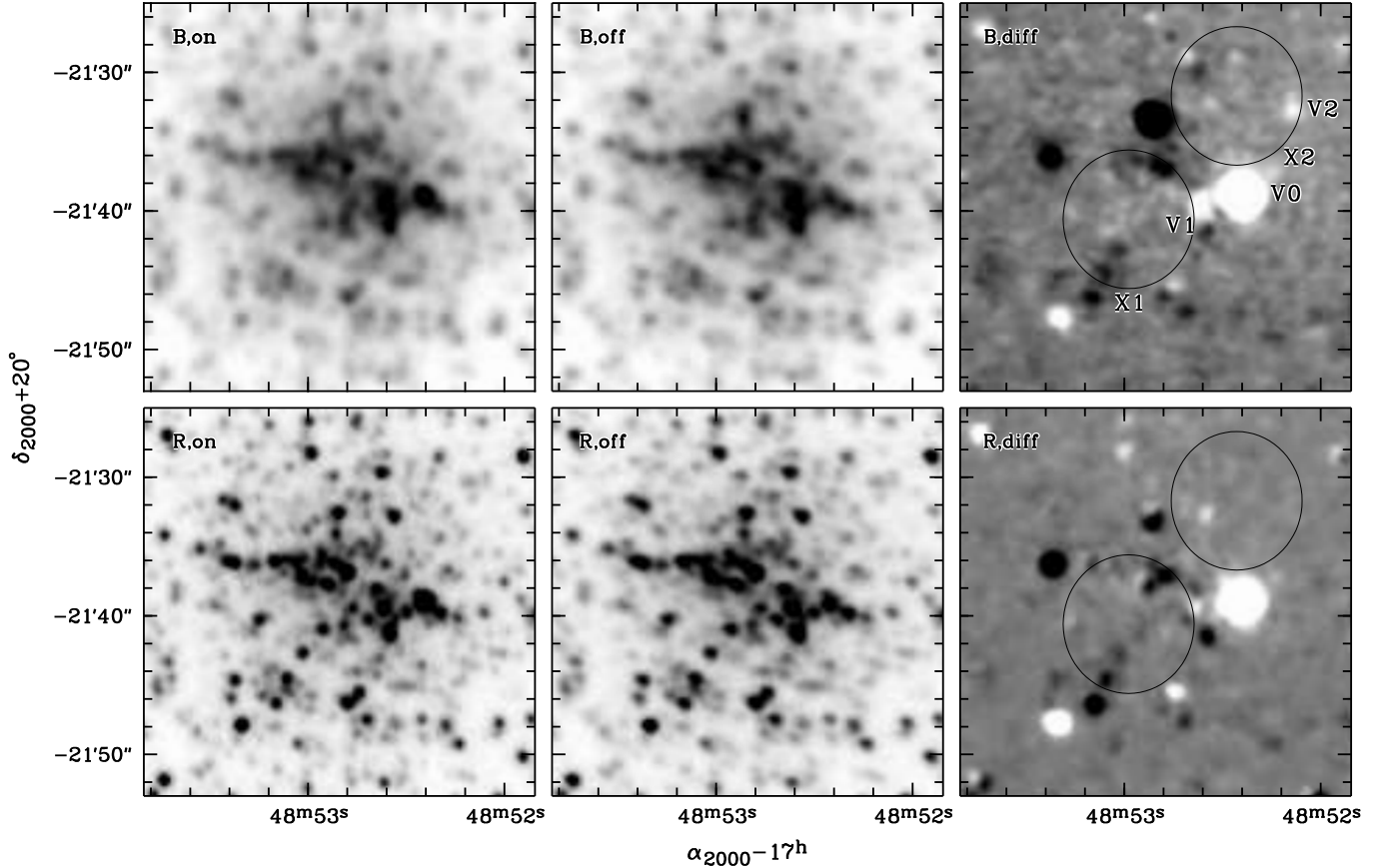,angle=-90,width=\textwidth}}
\caption[]{B and R images of the field of NGC~6440, taken with SUSI at
the NTT when the X-ray source was still on, and with FORS1 on Antu
when it was off.  
In these images, the brightest sources have
$B\simeq19$ and $R\simeq15$. Difference images are shown on the
right-hand side.  In these, stars that were brighter at the time the
X-ray source was on appear lighter, while those that were dimmer
appear darker.  The limiting magnitudes are $B\simeq24$ and
$R\simeq21$.  
The error circles mark the positions of the two X-ray
sources found in the ROSAT HRI images.  The radii are 5\arcsec,
approximately equal to the boresight uncertainty in the ROSAT
positions (the relative position between the two sources is
constrained much better; see text)
\label{fig:optical}}
\end{figure*} 

Optical images while the X-ray source was still on were taken for us
in Service Time in the night of 1998 August 26 to 27 at the 3.5-m New
Techology Telescope (NTT) at La Silla, using the Superb Seeing Imager
SUSI2, and with the 8-m Unit Telescope \#1 (Antu) of the Very Large
Telescope at Paranal, using the VLT Test Camera.  We will only discuss
the NTT observations here, as these had better seeing.  With SUSI2, one
10-s and two 100-s exposures were taken through a Bessell R filter, as
well as two 900-s exposures through a Bessell~B filter.  During the
observations, the seeing varied between $0\farcs7$ from the first
R-band image to $1\farcs2$ in last B-band image.  The night was not
photometric.  The detector was a mosaic of two EEV CCDs, each composed
of $2048\times4096$ square pixels of $15\,\mu$m on the side.  They
were read out binned by 2 in each direction, as the plate scale of
$0\farcs08{\rm\,pix^{-1}}$ would substantially oversample the seeing.
For all but the first, 10-s R-band image, the telescope was offset
such that the core of the cluster was not too close to the gap between
the two CCDs.  The data reduction was done using standard procedures,
determining the bias from the overscan regions (after verification on
bias frames) and correcting for pixel-to-pixel sensitivity variations
using dome flats taken in the morning following the observations.

On 1999 July 15, when the X-ray source was off, we took images with
the Focal Reducer/Low Dispersion Spectrograph FORS1 on Antu through
Bessell R, B, and U filters.  The night was not photometric, and the
seeing varied from $0\farcs8$ to $1\farcs2$.  Two 10-s and one 100-s
exposures were taken in~R, two 30-s and one 300-s in~B, and one 100-s
and one 600-s in~U.  The detector was a Tektronix CCD with
$2048\times2048$ pixels of $24\,\mu$m.  The standard resolution
collimator was used, for which the plate scale is
$0\farcs2{\rm\,pix^{-1}}$.  The detector was read out through all four
amplifiers, using the low-gain setting, of about
$3{\rm\,e^-\,ADU^{-1}}$.

The data reduction was done using standard procedures.  From bias
frames taken before and after the night, it was found that the level
was somewhat variable, both in time and in position on the detector,
but that the offsets remained constant relative to the levels found
from the overscan pixels (the latter determined separately for the
four quadrants).  For bias subtraction, therefore, we subtracted both
the levels from the overscan regions in individual frames, as well as
an average of the overscan-corrected bias frames.  The frames were
corrected for sensitivity variations using flat fields constructed from
images of the sky taken at dusk and dawn.

Since the conditions were not photometric during either of our two
runs, we cannot reliably calibrate our data.  We obtained a rough
calibration using B and R magnitudes of Hamuy (1986) of the
integrated flux of the cluster.  Magnitudes are listed for two
apertures, with diameters of 80 and 100\arcsec; these give consistent
results.  The calibration is consistent within 0.1\,mag with one
photoelectric B magnitude from Martins \& Harvel (1979;
star~5), within 0.3\,mag with B and R magnitudes inferred from V and I
magnitudes measured by Ortolani et al.\ (1994), but differs at
the 0.5 magnitude level from what one would infer using photographic B
magnitudes from Martins et al.\ (1980).  We therefore estimate
that the zero-point uncertainty on our quoted magnitudes is about
0.5\,mag.
\nocite{ham86}\nocite{mh79}\nocite{mhm80}

\subsection{Astrometry}

Astrometry of our frames was done relative to the USNO-A2.0 catalogue
(Monet et al.\ 1998).  \nocite{mbc+98} For all USNO-A2.0 stars overlapping
with the best-seeing FORS 10-s R-band image, centroids were determined
and the pixel coordinates corrected for instrumental distortion using
a cubic radial distortion function provided to us by T.~Szeifert and
W.~Seifert (1999, private communication).  From these, the zero point
position, the plate scale, and the position angle on the sky were
determined.  Unexpectedly, the root-mean-square residuals are rather
large, about $\sim\!1\arcsec$ in each coordinate.  This is
substantially larger than what we generally found in other fields; for
instance, for a field at $\alpha=18^{\rm{}h}56^{\rm{}m}$ and
$\delta=-38^{\circ}$, we find residuals of $0\farcs19$ in each
coordinate.

Most likely, the problem lies with the severe crowding in the field of
NGC~6440, which causes many of the USNO-A2.0 stars to be blended.
Furthermore, the USNO-A2.0 stars appear to have been derived from two
different sets of plates, with epochs 1950.465 and 1980.488.  The
former will be plates taken from Palomar, at high airmass.  For our
solution, therefore, we selected only measurements for the more recent
epoch, and from these, only those 246 stars which were well-exposed
and appeared stellar on our FORS image.  For these, the rms residuals
were $0\farcs53$ and $0\farcs59$ in right ascension and declination,
respectively.  This is still much larger than usual, and therefore it
is difficult to estimate the uncertainty. 
Fortunately, there is one
Tycho star, TYC 6257-368-1, at the Southern edge of our VLT images.
While this star is strongly overexposed, it was possible to determine
a reasonably accurate centroid (verified using different exposures).
The coordinates we infer using our astrometry are offset from the
Tycho-2 coordinates (H\o{}g et al.\ 2000) by $-0\farcs27$ and
$-0\farcs14$ in right ascension and declination, respectively.  We
conclude that the systematic uncertainty in our astrometry is
$\sim\!0\farcs3$, i.e., substantially less than the uncertainty in the
X-ray positions.  \nocite{hfm+00}

The individual frames were tied in to the astrometry using some 400
secondary reference stars in a square region 2\arcmin\ on a side
centered on NGC~6440, but excluding stars within 20\arcsec\ of the
core.  Typical residuals range from $0\farcs016$ in each coordinate
for the tie to the SUSI R-band images to $0\farcs08$ for that to the
FORS U-band images.

\subsection{Variable sources}

In Fig.\,\ref{fig:optical}, the reduced on and off-state B and R-band
images are shown.  The positions of the two X-ray sources fall close
to the core of NGC~6440 and because of the crowding it is difficult to
discern variable stars directly.  To get an objective measure of
variability in the core, we formed on-off difference images.  For this
purpose, we used the optimal image subtraction technique introduced by
Alard \& Lupton (1998).  In this method, the image with better
seeing is convolved with a kernel chosen such that the convolved
point-spread function is as close as possible to that of the image
with the worse seeing.  In our case, this did not work perfectly,
because for both bands the seeing is comparable in the on and off
images, while the shapes of the point spread functions 
are somewhat different; 
the result is a convolution kernel which
is negative in some points (in directions where ``sharpening'' of the 
better-seeing image is required).  
We proceeded by convolving the worst-seeing (FORS) images with a
Gaussian with $\sigma=0\farcs2$.  Relative to this slightly smoothed
image, it was possible to determine a kernel which is positive
everywhere.  After convolution with this kernel, the SUSI image and
the slightly smoothed FORS image have very similar point-spread
functions, and the subtraction works well.

The on minus off difference images found using the above method are
shown in Fig.\,\ref{fig:optical}.  It is clear that a number of stars
brightened or dimmed considerably; a conspicuous one is star V0 on
the Western end of the core.  Most of these variables,
including V0, are bright, red
stars, most likely near the end of the asymptotic giant branch;
indeed, the reddest, most luminous object in the sample of Ortolani
et al. (1994) was found to be variable as well (it is out of
the field shown in Fig.\,\ref{fig:optical}).  \nocite{obb94}

If the optical emission of the X-ray source is dominated by an
accretion disk, as is generally the case for low-mass X-ray binaries,
the optical counterpart is expected to be blue, and the brightess
difference between on and off is expected to be relatively large.
Given the problems with crowding, the resolution element that includes
the source may not appear to be blue, but the difference between on
and off should be blue.  
For this reason, we do not consider V0 to be a likely candidate;
more likely, it is a Mira type variable. One could envisage a wide binary
in which a Mira star transfers matter during its expansion. However,
in that case the outburst would be expected to last rather longer
than observed.

In each of the error circles, there is one source which was brighter
when the X-ray source was on, and for which the difference flux is
substantially bluer than that for other variables in the field.  For
X1, the source V1 is just outside the western edge of the error circle, 
while for X2, V2 is just inside the error circle. The coordinates of
these variables are given in Table\,\ref{tabpos}.V2 may be interesting in
particular, since the relative brightening for that source is the
second-largest in the field (about 10\% in B, $<\!4\%$ in R; the
largest relative brightening -- 50\% and 100\%, respectively -- is
shown by V0).  

If V1 or V2 were the optical counterpart of the X-ray transient, we would 
expect
that the source in quiescence had negligible contribution to the optical flux
in the off image. The excess flux in the on image then would correspond
to the the total flux of the source in outburst.
We estimate the corresponding magnitudes using our approximate calibration.
For V1 we find $B\simeq22.0$, $R\simeq19.3$, and for V2
$B\simeq22.7$, $R>21.1$.  Correcting for
reddening, one infers $(B-R)_0\simeq1.1$ and $(B-R)_0\la0$,
respectively, again indicating that V2 is very blue.

From the above, it seems that V2 has all the characteristics expected
for the optical counterpart.  It is clear, however, that finding a
variable source inside one of the two error circles is not unlikely,
even one whose variation is relatively blue.  Therefore, at present we
consider this source as no more than an interesting candidate.

\section{Discussion}

\begin{figure}
\centerline{\psfig{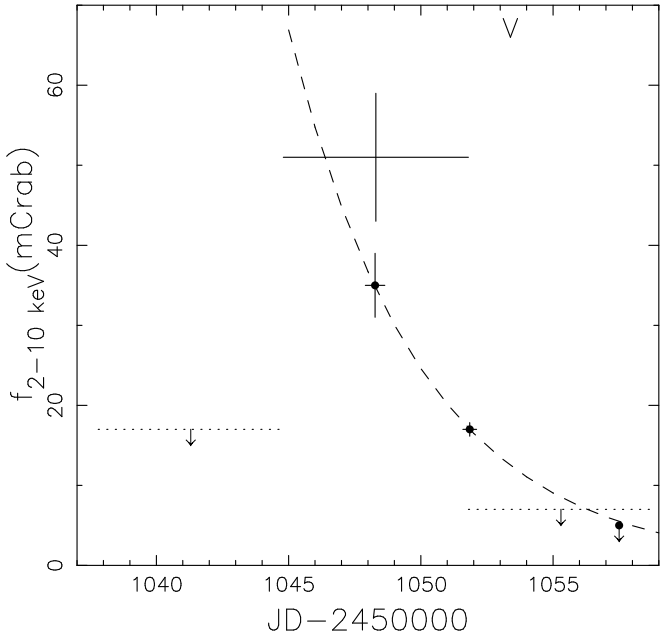} {\hfil}}

\caption{The August 1998 outburst of the X-ray transient in 
NGC\,6440 as observed
with BeppoSAX  ($\bullet$), and with the XTE All Sky Monitor (horizontal 
lines, solid for the detection, dotted for upper limits). 
The dashed line indicates exponential decay with e-folding time 5 days, 
passing through the BeppoSAX detections.
The V indicates the time of the optical observations.
\label{xcur}}
\end{figure}

We have tried to reconstruct the approximate X-ray light\-curve of the
August 1998 outburst by combining data from various measurements.
Detections were made by the Wide Field Cameras and the Narrow Field
Instruments on board of BeppoSAX on Aug 22 and 26, respectively;
and by the XTE All Sky Monitor in a seven-day period starting
on Aug 19. We revise the flux detected with the Wide Field Cameras
to $35\pm4$\,mCrab, slightly upwards from the values given in In 't Zand
et al.\ (1999), on the basis of a better calibration.
Upper limits were obtained with the BeppoSAX Wide Field Cameras on Sep 1
and with the XTE All Sky Monitor in the seven days periods preceding and
following the detection.
Fig.\,\ref{xcur} shows the resulting lightcurve.

Extrapolating the 5-day exponential decay from the BeppoSAX detections to the
time of the ROSAT HRI observation, we predict an X-ray luminosity of
$\sim 10^{35}\ergs$ in the ROSAT band; this is about an order of magnitude
above the observed upper limit.
This may imply that the decay accelerated; or alternatively that the
spectrum softened, since low-energy photons are much more affected by the
heavy absorption towards NGC\,6440.
There is indeed evidence for other X-ray transients that the spectrum
in the low state is much softer than during outburst, e.g.\ for Aql X-1
(Verbunt et al.\ 1994).\nocite{vbj+94}

The X-ray lightcurve shown in Fig.\,\ref{xcur}
implies that the optical observations were made at
an X-ray flux level of about 10\,mCrab, corresponding to a source
luminosity $\simeq2\times10^{36}\ergs$.
Van Paradijs \&\ McClintock (1994) give a semi-empirical relation between
the orbital period, X-ray luminosity and absolute visual magnitude
of a low-mass X-ray binary. Applying this relation to the transient
in NGC\,6440 with the estimate of the X-ray luminosity for the time of the
optical observation, we obtain $M_{\rm V}\simeq4.0$ for an assumed
1\,hr period. At the distance and reddening of NGC\,6640 this corresponds
to $V\simeq21.7$. \nocite{vpm94}
The intrinsic $B-V$ colour of low-mass X-ray binaries is close to zero; 
with the reddening to NGC\,6440 we thus predict $B\simeq22.7$ for a one hour 
period.
For a period of 5\,hr (0.2\,hr) the predicted magnitude is about 1 magnitude
brighter (fainter).
We conclude that the candidate in the error circle of X2 is viable;
the proximity of the predicted $B$ magnitude to the observed one is fortuitous,
considering that the spread in the relation given by Van Paradijs \&\
McClintock is about a magnitude, 
and that our estimate of the X-ray luminosity is uncertain.
We therefore dare not estimate an orbital period on the basis
of the magnitude of our candidate.

The core of NGC\,6440 contains PSR\,B\,1745$-$20 (Lyne et al.\ 1996).
The total energy loss $\dot E\equiv I\Omega\dot\Omega$ for the pulsar
is about $6.6\times 10^{32}\ergs$, where 
$I$ is the moment of inertia of the neutron star, $\Omega$ its
rotation frequency and $\dot\Omega$ the time derivative of $\Omega$.
Typical X-ray luminosities for radio pulsars are of order
$L_{\rm x}\sim 10^{-3}I\Omega\dot\Omega$ 
(e.g.\ Fig.\,4 in Verbunt et al.\ 1996).\nocite{vkb+96}
We conclude that it is very unlikely that the pulsar is responsible
for the observed X-ray flux of X1 or X2.

\begin{acknowledgements}
We have made use of the ROSAT Data Archive of the Max Planck
Institut f\"ur extraterrestrische Physik at Garching; and of the SIMBAD 
database operated at Centre de Donn\'ees astronomiques in Strasbourg.
The ROSAT HRI data of 8 Sep 1998 and the ESO
data of 26/27 Aug 1998 were obtained as Target of Opportunity
observations, triggered by quick look analysis of BeppoSAX observations.
We thank Michael Smith for making the BeppoSAX
observations quickly available, and  Vanessa Doublier and Bruno Leibundgut
for making the ESO NTT and VLT observations.
\end{acknowledgements}

\end{document}